\documentclass[reprint,amsmath,amssymb]{revtex4-1}
\usepackage[utf8x]{inputenc}
\usepackage{graphicx}
\usepackage{dcolumn}
\usepackage{bm}
\usepackage{color} 
\usepackage{amsmath}
\usepackage{tensor}
\usepackage[english]{babel}
\usepackage{amsthm}
\usepackage{amssymb}

\usepackage{soul} 
\usepackage[normalem]{ulem} 

\begin{document}

\preprint{APS/123-QED}


\title{A numerical study of the localization transition of  Aubry-Andr\'{e} type models}

\author{Bal\'azs Het\'enyi$^{1,2}$ and Istv\'an Balogh$^1$}
\affiliation{$^1$MTA-BME Quantum Dynamics and Correlations Research Group, Department of Physics, Budapest University of Technology and
  Economics, H-1111 Budapest, Hungary \\ and \\
 $^2$Institute for Solid State Physics and Optics, Wigner Research Centre for Physics,  H-1525 Budapest, P. O. Box 49, Hungary}

\date{\today}

\begin{abstract}
We use tools based on the modern theory of polarization for a numerical study of the localization transition of the Aubry-Andr\'{e} model.   In this model the spatial modulation of the potential, $\alpha$, is an irrational number, which we approximate as the ratio of Fibonacci numbers, $F_{n+1}/F_n$, where $F_n=L$ is also the system size.  We calculate the phase diagram as a function of particle density (filling) and potential strength $W$.  We calculate the geometric Binder cumulant and also apply a renormalization approach.   At any given finite system size we find that at many densities the transition occurs at or near $W=2t$ ($t$ denotes the hopping).   This is where single particle states are known to localize.   However, we also find "spikes", densites at which the transition occurs in the range $0<W<2t$.  These spikes occur for densities at which there are no partially filled bands.  As the system size (and both $F_n$ and $F_{n+1}$ in $\alpha$) is increased  the spikes tend towards zero, but the density at which they occur also changes slightly: they approach irrational numbers which can be written as Fibonacci ratios or sums thereof.  For densities which are fixed ratios for all system sizes, the transition occurs at $W=2t$.  We also study an extension of the original Aubry-Andr\'{e} model with second nearest neighbor hoppings.  This model also exhibits a distorted phase diagram compared to the original one, with spikes which do not necessarily tend to zero, but to finite values of $W$, determined by the modifed gap structure.
\end{abstract}


\maketitle

\section{Introduction}   

The delocalization-localization transition (DLT) occuring in disordered~\cite{Abrahams79,Langedijk09,Evers08} and quasiperiodic~\cite{Aubry80,Martinez18,Dominguez-Castro19} systems is a topic of ongoing interest.   Recent experimental developments allow the manipulation of ultracold atoms in optical lattices and enable direct examination.  Billy et al. investigated~\cite{Billy08} Anderson localization in a Bose-Einstein condensate (BEC), Roati et al. observed~\cite{Roati08,Modugno09} the duality~\cite{Johansson91} of the Aubry-Andr\'{e} model~\cite{Aubry80} (AAM) in a BEC in a quasiperiodic optical lattice.   The AAM is equivalent to the Harper model~\cite{Harper55}, which describes quantum Hall systems~\cite{vonKlitzing80,Tsui82,Thouless82}.  Extended~\cite{Biddle10,Biddle11,Ganeshan13,Ganeshan15,Bistritzer11,Monthus17,Li20,Padhan22,Goncalves23a,Dziurawiec24} AAMs exhibit an even richer variety of physics, including mobility edges~\cite{Biddle10,Biddle11,Li20}, multiple localization transitions~\cite{Padhan22} and multifractality~\cite{Monthus17,Dziurawiec24}.  Recently, twisted bilayer graphene~\cite{Bistritzer11}, a two-dimensional material was studied~\cite{Goncalves23a} using  an extended AAM derived by dimensional reduction to one dimension.  The phase diagram was calculated using a renormalization group (RG) approach~\cite{Goncalves23b}.   Studies~\cite{Papp07,Dey20,Ganguly23} of small AAM rings by Dey et al.~\cite{Dey20} suggest that they can be used as high-to-low conductivity switches by varying the filling.  Similar behavior was demonstrated~\cite{Ganguly23} in an AAM assembled from electric circuit elements.\\

The crucial aspect of the AAM, which sets it apart from most condensed matter models, is that the modulation depends on an irrational parameter $\alpha$, often taken to be the golden ratio.  This aspect motivated many mathematical studies~\cite{Jitomirskaya99,Avila09,Avila23}, which focused on the nature of the band structure and the characteristics of single particle states.  Jitomirskaya~\cite{Jitomirskaya99} proved that such states localize at a finite value of the interaction strength ($W=2t$, where $t$ denotes the hopping parameter).   Avila et al.~\cite{Avila09} showed that the single-particle energy spectrum of the AAM  is a Cantor set.   A very recent study~\cite{Avila23} showed that commensurate systems exhibit gaps (when $\alpha$ is approximated as a rational number) which survive in the incommensurate limit (when the irrational limit for $\alpha$ is taken).   In mathematical circles, for historical reasons, this set of problems is referred to as the "Ten Martini Problem"~\cite{Jitomirskaya99,Avila09,Avila23}.  Some studies argue~\cite{Wang17,Zhang15} for refining these results in some special circumstances.\\

In this work we calculate the phase diagram of the AAM as a function of particle density (filling) and the normalized parameter potential strength $W/t$.  All our calculations are based on the many-body ground state wave function, which is a Slater determinant of single particle eigenstates.    We study the model under periodic boundary conditions (PBC) for finite system sizes based on a rational approximation to $\alpha$ as the ratio of Fibonacci numbers $F_{n+1}/F_n$, where $F_n=L$, the system size, but we extrapolate to the irrational limit by comparing different system sizes.  To gauge localization and to determine the DLT we use the geometric~\cite{Hetenyi19,Hetenyi22,Hetenyi24} Binder cumulant~\cite{Binder81a,Binder81b} (GBC), a tool rooted in the modern polarization theory~\cite{King-Smith93,Resta94,Vanderbilt18,Resta00} (MPT).  In addition, we also carry out a position space renormalization~\cite{Hetenyi21} calculation, also rooted in MPT.  We will refer to this second method as polarization amplitude renormalization (PAR).  \\

Our main findings are as follows.  For any given system size, we find that as a function of particle density there are fillings at which the DLT occurs at or near $W=2t$, which is where single particle states are known~\cite{Aubry80,Jitomirskaya99} to localize.  We also find "spikes" in the phase diagram, by which we mean densities at which the critical $W$ is in the range $0<W<2t$, and significantly smaller than $2t$.   Upon increasing the system sizes, and corresondingly increasing $F_{n+1}$ and $F_n$ in the approximate $\alpha$, the spikes "grow", meaning they get closer to zero (new spikes also appear).  The "grown" spikes are not exactly at the same densities, but we show that as the system size increases the densities for the spikes tend to a particular set of irrational numbers.  The characteristic of this set of irrational numbers is that they can be written as the ratios of Fibonacci numbers, $F_m/F_n$, where $m<n$, or sums thereof.  From this extrapolation procedure we conclude that in the thermodynamic limit, fillings corresponding to irrational numbers to which Fibonacci ratios tend are fully localized, and no DLT occurs at finite $W$.  Inspecting the band structures that correspond to the different system sizes, we find that at these fillings there are no partially filled bands.  It is an interesting example of Kohn's tenet~\cite{Kohn64,Vanderbilt18,Resta00} which states that whether a system is conducting or not depends ultimately on localization of the many-body wave function, rather than on the localization of single particles, as one would expect classically.   We also carry out a study on an extended version of the AAM, with second nearest neighbor hoppings.   We find similar spikes, but they do not necessarily tend to $W=0$, but instead to finite values of $W$, determined by the modified gap structure of the extended model.  For the original model, with nearest neighbor hoppings only, gaps open at $W=0$ and they close at $W \rightarrow \infty$.  Turning on a second nearest neighbor hopping distorts the band structure, and gaps open at finite values of $W$.  These gap openings function as lower limits to the critical $W$ at the relevant fillings.\\
  
 \begin{figure}[ht]
 \centering
 \includegraphics[width=7cm,keepaspectratio=true]{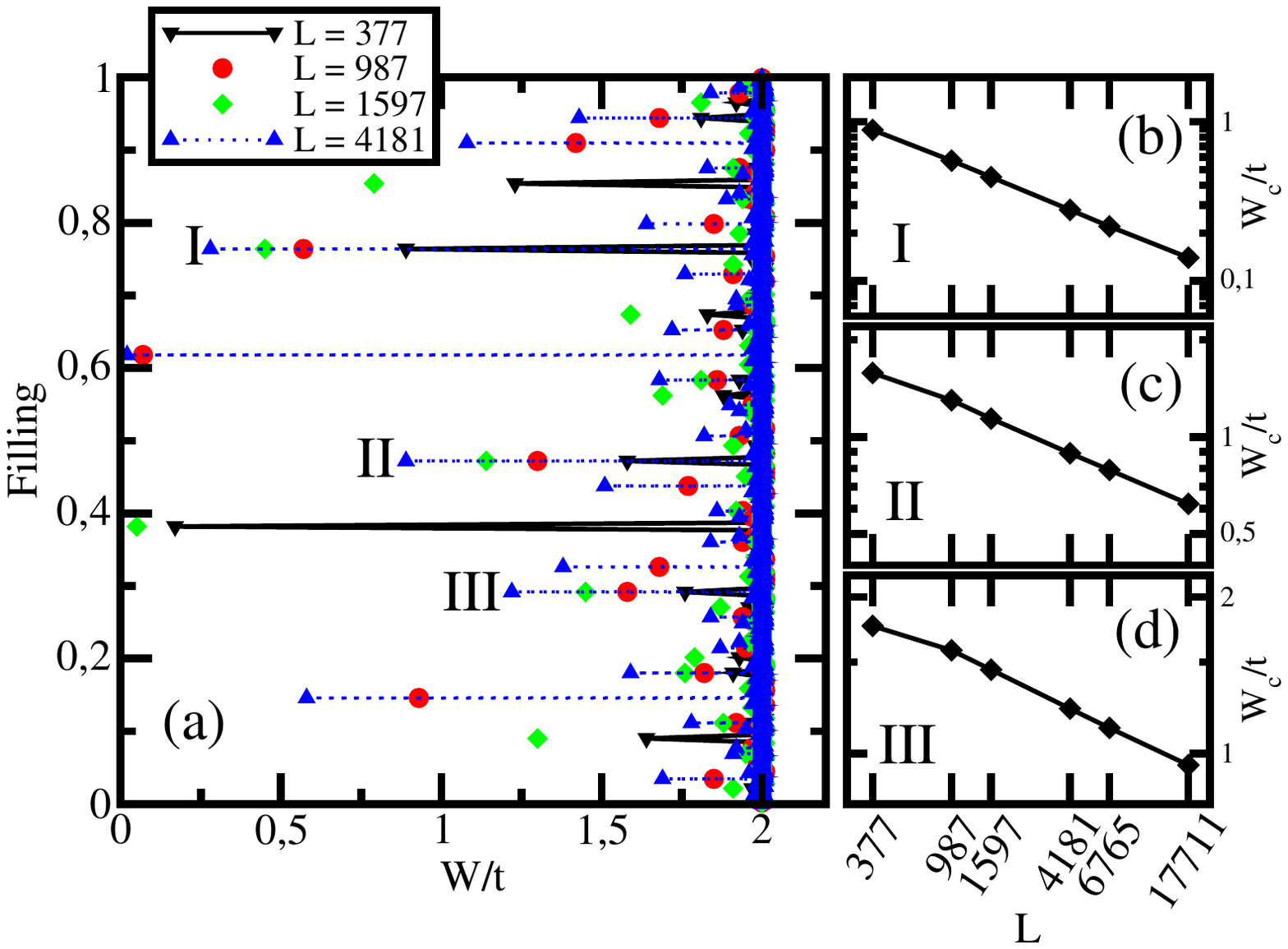}
 \caption{Panel (a) on the left shows the calculated phase diagram for four different system sizes, $L=377,987,1597,4181$.  The predicted $W/t=2$ line is found for most fillings, but all system sizes exhibit spikes where the localization transition occurs at $W/t<2$.  The right panels ((b), (c), and (d)) show the critical $W$ for three spikes as a function of system size for six different sizes, $L=377,987,1957,4181,6765,17711$.  The spikes are labeled in panel (a) as I, II,  and III.  In the plots for the three spikes, (b), (c), and (d), the scales of all axes are logarithmic.  }
 \label{fig:pdfill}
\end{figure}

This paper is organized as follows.  In the next section, we present the models to be studied.  In section \ref{sec:method}, we present the two methods used here, the GBC and the PAR.  In this section we also present the essence of our extrapolation (how extrapolation to rational vs. irrational fillings are done).  This uses the Zekendorf construction, which relates natural numbers to the Fibonacci sequence.  The construction itself is presented in an appendix.   In section \ref{sec:results} we present our results.  In section \ref{sec:cnclsn} we conclude our work.

\section{Aubry-Andr\'{e} Model and extension to second nearest neighbor hopping} 

We consider a Hamiltonian of the Aubry-Andr\'{e} type given by
\begin{equation}
\label{eqn:HAA}
\hat{H}_{AA} = -t \sum_{j=1}^L (c_{j+1}^\dagger c_j + c_j^\dagger c_{j+1}) + W \sum_{j=1}^L \xi_j n_j,  
\end{equation}
where $\xi_j = \cos \left( 2 \pi \alpha j \right)$, where $\alpha$ is the golden ratio obtained from the ratio of consecutive members of the Fibonacci sequence in the limiting case, and the operators $c_j^\dagger$($c_j$) create(annihilate) a particle at site $j$.   For finite systems with PBC, we approximate the irrational $\alpha$ as the ratio $\alpha \approx F_{n+1}/F_n$, where $F_{n}$ is the $n$th Fibonacci number, so the size of the system can not be smaller than $F_n$, otherwise the potential would not be "smooth".  In our calculations $L=F_n$. \\ 

As mentioned in the introduction, extensions of the Aubry-Andr\'{e} model constitute a very broad topic.  Here we treat a simple extension to test the robustness of our results, namely the model defined in Eq. (\ref{eqn:HAA}) with an addtional term consisting of second nearest neighbor hopping,
\begin{equation}
\label{eqn:Hext}
\hat{H}_{ext} = \hat{H}_{AA}-t_2 \sum_{j=1}^L (c_{j+2}^\dagger c_j + c_j^\dagger c_{j+2}),
\end{equation}
again, with PBC.  Our objective is to further test the robustness of our results.\\

\section{Methods}
\label{sec:method}

In this section the two methods used in this work are described; the geometric Binder cumulant, which is the MPT implementation of the Binder cumulant developed for locating phase transition points via finite size scaling, and a position space based renormalization approach also based on the MPT.   Our results are based on extrapolations to rational vs. irrational fillings in the thermodynamic limit.  After the two numerical methods we also explain how our extrapolations work, using the Zekendorf construction as a justification.\\

\subsection{The geometric Binder cumulant}

In this study we apply PBC for the AAM.  For systems with PBC the Bloch theorem holds, however, quantities which gauge localization are not immediate, because the position operator is ill-defined.  This difficulty is overcome by the MPT~\cite{King-Smith93,Resta94,Vanderbilt18,Resta00} which furnishes cumulants sensitive to localization.   In topological insulators, topological invariants~\cite{Bernevig13,Asboth16,Thouless82,Kitaev01} are defined under PBC.  The geometric Binder cumulant was introduced in Refs. \cite{Hetenyi19,Hetenyi22} and was applied to the Aubry-Andr\'{e} model at half filling in Ref. \cite{Hetenyi24}.  Our application here follows this last reference.    \\

We diagonalize $\hat{H}_{AA}$ under  PBC, meaning that we obtain a set of states on the lattice, $\Phi_\lambda(j)$, where $\lambda$ denotes the state index, and $j$ denotes the lattice site.  For a system with $N$ particles, the ground state wave function is a Slater determinant of the $N$ lowest energy states,
\begin{equation}
\Psi(j_1,...,j_N) = \mbox{Det} \left[ \Phi_\lambda(j_\mu) \right]; \lambda = 1,...,N.
\end{equation}
To construct the GBC, one first calculates the quantity $Z_q$ (sometimes known as the polarization amplitude~\cite{Kobayashi18}), defined in the case of  band systems as
\begin{equation}
Z_q  =  \mbox{Det} \left[ U_{\lambda \lambda'}^{(q)} \right],
\end{equation}
where
\begin{equation}
U_{\lambda \lambda'}^{(q)}= \sum_{j=1}^L \phi^*_\lambda(j) \exp\left( i \frac{2 \pi q}{L}j\right) \phi_{\lambda'}(j).
\end{equation}

$Z_q$, the quantitiy known as the polarization amplitude,  is really the cumulant generating function of the distribution of the total position $X$ of the charge carriers of the system, $P(X)$.   $P(X)$ can be obtained from the inverse discrete Fourier transform of $Z_q$,
\begin{equation}
\label{eqn:PX}
P(X) = \frac{1}{\sqrt{L}} \sum_{q = 0}^{L-1} Z_q \exp \left( i \frac{2 \pi q}{L}X\right).
\end{equation} 
$Z_q$ is different from usual generating functions often seen in statistics in one aspect: because of the periodic boundary conditions, $Z_q$ is defined on a discrete set of points ($q=1,...,L$), rather than on the continuum.   Because of this, one can only take discrete derivatives with respect to $q$. \\

Our first step is to take the absolute value of all $Z_q$.  This step is equivalent to centering the distribution $P(X)$.  Using these centered distributions, we can define the approximate second and fourth order statistical moments as
\begin{eqnarray}
\label{eqn:M2M4}
M_2 &=& \frac{L^2}{2 \pi^2} (1 - |Z_1|)\\
M_4 &=& \frac{L^4}{8 \pi^4} \left( - |Z_2| + 4 |Z_1| - 3\right), \nonumber
\end{eqnarray}
From which we can define the GBC as
\begin{equation}
\label{eqn:U4}
U_4 = 1 - \frac{1}{3} \frac{M_4}{M_2^2}.
\end{equation}
$U_4$, written as in Eq. (\ref{eqn:U4}), has the usual form of the Binder cumulant~\cite{Binder81a,Binder81b} (where $M_4$, $M_2$ indicate usual statistical moments of some order parameter), and is a known gauge of phase transitions.  In the MPT context, the moments, (Eq. (\ref{eqn:M2M4})) do not take usual forms (operator expectation values), because the relevant observable in this case is cast as a geometric phase.   The GBC ($U_4$) changes sign at a critical $W/t$, as was shown in Ref. \cite{Hetenyi24}. \\

\subsection{Polarization Amplitude Renormalization}

It is possible to apply RG methods~\cite{Goncalves23b,Wilkinson84,Niu86,Ashraff88,Andrews24} to quasi-periodic systems.  One such approach~\cite{Goncalves23b} is based on increasing the unit cell of commensurate approximants.  In a recent work an MPT based  renormalization group approach was developed~\cite{Hetenyi21}. We can extend this method in such a way that rational and irrational fillings can be distinguished.   The idea is to apply decimation to the expectation value of the twist operator, $Z_1$.  The approach generates a flow in the parameters of the Hamiltonian.  The flow equations can exhibit fixed points (repulsive or attractive).   In our case, one can generate a flow in $W$ by iterating the following equation,
\begin{equation}
\label{eqn:Z1RG}
|Z_1(W_{j+1},N',L')| = |Z_1(W_j,N,L)|,
\end{equation}
where the approach of Ref. \cite{Hetenyi21} was modified by including the particle number $N$ as a variable.  In practice, this equation translates to the following: one starts with a value of the potential strength, $W_0$, and calculates $Z_1$ for a system with  particle number $N$, and system size $L$.  One then finds the $W_1$ that gives the same value of $Z_1$ for a smaller system with particle number $N'$ and size $L'$.  In the next step, one uses $W_1$ again, to calculate $Z_1$ for the $N,L$-system.  In this case the iteration generates a flow in $W$.  It is possible to distinguish rational and irrational fillings.  For a rational filling, one keeps the particle number so that $N/L$ is fixed during the iteration.  For irrational filling one has to make the ratios $N/L$ and $N'/L'$ correspond to different approximations of the desired irrational filling, the $N'/L'$ being the one closer.\\ 

\subsection{Remarks on the Fibonacci sequence: irrational and rational filling limits}

In this subsection we describe certain aspects of the Fibonacci sequence necessary for the analysis of our results.   For further information, we suggests Refs. \cite{Vajda89,Samons94}.  In the Appendix we outline the Zekendorf construction, which casts any natural number as a sum of Fibonacci numbers.  \\

The Fibonacci sequence is generated by the recursion relation,
\begin{equation}
F_{n+1} = F_n + F_{n-1},
\end{equation}
and by taking as initial values, $F_0=1$ and $F_1=1$.  In the limit $n\rightarrow \infty$ the ratio $F_{n+1}/F_n$ tends to the number known as the golden ratio.  This result can be obtained by setting $x = F_{n+1}/F_n$ and solving the quadratic equation which corresponds to the $n\rightarrow \infty$ limit of the recursion relation,
\begin{equation}
x^2 + x - 1 = 0.
\end{equation}
This equation gives two roots,
\begin{equation}
\label{eqn:xpm}
x_{\pm} = \frac{1 \pm \sqrt{5}}{2},
\end{equation}
and $x_+$ corresponds to the golden ratio ($x_+=\alpha$).  With the help of $x_{\pm}$ one can write a Fibonacci number as 
\begin{equation}
F_n = \frac{x_+^n - x_-^n}{x_+ - x_-}.
\end{equation}

Zekendorf has constructed~\cite{Vajda89,Samons94} a unique representation of all natural numbers by Fibonacci sequences.  Any natural number, $N$ can be written as the sum of distinct Fibonacci numbers.  For uniqueness one further restriction is needed: there should be no consecutive Fibonacci numbers appearing in the sum.  For some natural number, $N$, we can write a general expression of the form,
\begin{equation}
N = \sum_{w=1}^k F_{m_w}.
\end{equation}
which, since the system size itself is a Fibonacci number, leads to a result for the filling,
\begin{equation}
\frac{N}{L} =  \sum_{w=1}^k \frac{F_{m_w}}{F_n} < 1.
\end{equation}
The ratios $F_{m_w}/F_n$ can be expressed as
\begin{equation}
\label{eqn:FperF}
 \frac{F_{m_w}}{F_n} = \frac{x_+^{m_w} - x_-^{m_w}}{x_+^n - x_-^n}.
\end{equation}
Increasing system size corresponds to shifting all indices, $m_w$ as well as $n$, by the same integer.  For some larger system, we can write the filling as,
\begin{equation}
\frac{N}{L} =  \sum_{w=1}^k \frac{F_{m_w+s}}{F_{n+s}},
\end{equation}
for some integer $s>0$.   As $s \rightarrow \infty$ each ratio, $F_{m_w+s}/F_{n+s}$, tends to a known irrational number (can be calculated from knowing the roots $x_+$ and $x_-$ and using Eq. (\ref{eqn:FperF})), therefore the sum itself, $N/L$, will also tend to some known irrational number as $L \rightarrow \infty$.   This is our procedure for extrapolating to irrational fillings. \\

On the other hand, it is also possible to keep $N/L$ a fixed ratio when the thermodynamic limit is taken.   This is, of course, standard procedure in most finite size scaling calculations.  In this case, for different system sizes, the Fibonacci numbers entering the sum for $N$ will be entirely different sums, and the above argument is no longer valid.  \\

Let us consider two simple examples to illustrate this point.   First we consider an extrapolation to an irrational filling.  Suppose that we start with a system of size $L=21$, and we set the number of particles to be $N=11$.  This corresponds to a filling of $N/L = 11/21 \approx 0.523...$.  Using the Zekendorf construction to write $N$ as a sum of nonconsecutive Fibonacci numbers, we obtain,
\begin{equation}
\frac{N}{L} = \frac{8+3}{21} = \frac{F_4 + F_6}{F_8}.
\end{equation}
To extrapolate to an irrational number in the thermodynamic limit, we can shift all indices by the same integer.  Using a slight relabeling, 
\begin{equation}
\frac{N}{L} = \frac{F_{n-4} + F_{n-2}}{F_{n}} 
\end{equation}
As $n\rightarrow \infty$ $N/L \rightarrow 5 - 2 \sqrt{5} = .527864.....$, as can be shown by substituting $x_+$ and $x_-$ from Eq. (\ref{eqn:xpm}) into Eq. (\ref{eqn:FperF}).  \\

If one is to study a rational filling, in that case, as the system size is increased, the ratio $N/L$ is kept the same.  For a simple example, let us consider the filling $N/L = 1/3$.  To realize this, $L$ needs to be divisible by three.  As a first example, we choose the Fibonacci number $L=21$, for which $N=7$.  Using the Zekendorf construction, $N=7$ can be written,
\begin{equation}
N = F_3 + F_5 = 2 + 5.
\end{equation}
The next possible $L$ is $144$, for which $N=48$.  Applying, again, the Zekendorf construction,
\begin{equation}
N = F_1 + F_7 + F_9 = 1 + 13 + 34.
\end{equation}
In this case, the two sums (one for $N=7$, the other for $N=48$) do not even contain the same number of terms.   When extrapolation to a rational filling is done by increasing system size, the sum representing $N$ corresponds to sums of different types, which can not be represented iteratively as a simple shift of indices.    \\

In this sense there is a qualitative difference between the irrational versus the rational approximation, which, as shown below, leads to quantitatively different results for the DLT. \\ 

\section{Results and Analysis}

\label{sec:results}

In this section we present the results of our calculations.  In the first subsection, we will show the band structure of the system, scaled in such a way, that the duality is immediately obvious.  We then address a technical issue about odd vs. even particle number/system size.  Our main results are the GBC based results for the phase diagram (subsection \ref{ssec:GBC}) and its further corroboration via the PAR method (\ref{ssec:PAR}).  The robustness of our results is further supported by the phase diagram of the extended AAM. 

\subsection{Energy levels} 

\begin{figure}[ht]
 \centering
 \includegraphics[width=7cm,keepaspectratio=true]{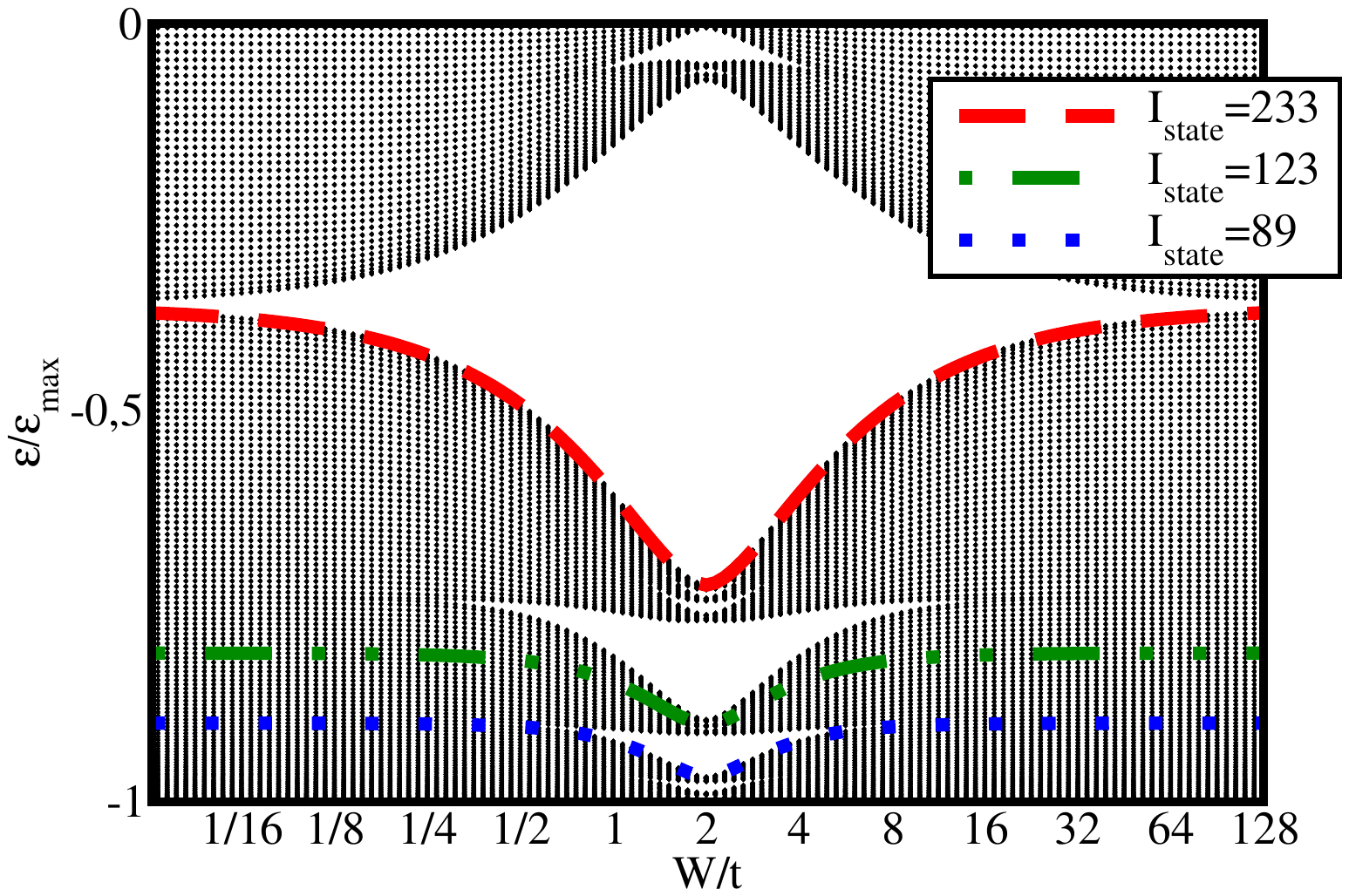}
 \caption{Energy levels of the Aubry-Andr\'{e} model for a system of size $L=610$ with periodic boundary conditions as a function of $W/t$.  Only states with $\epsilon<0$ are shown, since the system is symmetric with respect inversion around $\epsilon=0$.  The scale of the $W/t$ axis is logarithmic.   Band gaps are found at states whose indices correspond to Fibonacci numbers or sums of Fibonacci numbers. }
 \label{fig:band_structure}
\end{figure}

Fig. \ref{fig:band_structure} shows the energy levels as a function of $W/t$ for $L=610$.  Only states with energies below zero are shown, since the upper half is symmetric to the lower half.  The $W/t$-axis is logarithmic, which makes the other symmetry, originating from duality explicit (symmetry around $W=2t$).   As
 $W$ is increased from zero, gaps are seen to open.  In Fig. \ref{fig:band_structure} the uppermost states for three filled bands are indicated.  The state indices are at $I_{state} = 89, 123,233$, two of which are Fibonacci numbers, the remaining one is the sum of two Fibonacci numbers.  Not all gaps are visible in the figure, others can be seen by zooming in, but once that is done, all gaps start from $W/t=0$ and close again at $W/t \rightarrow \infty$.\\

\subsection{Even vs. odd $N$ and $L$}  

\begin{figure}[ht]
 \centering
 \includegraphics[width=7cm,keepaspectratio=true]{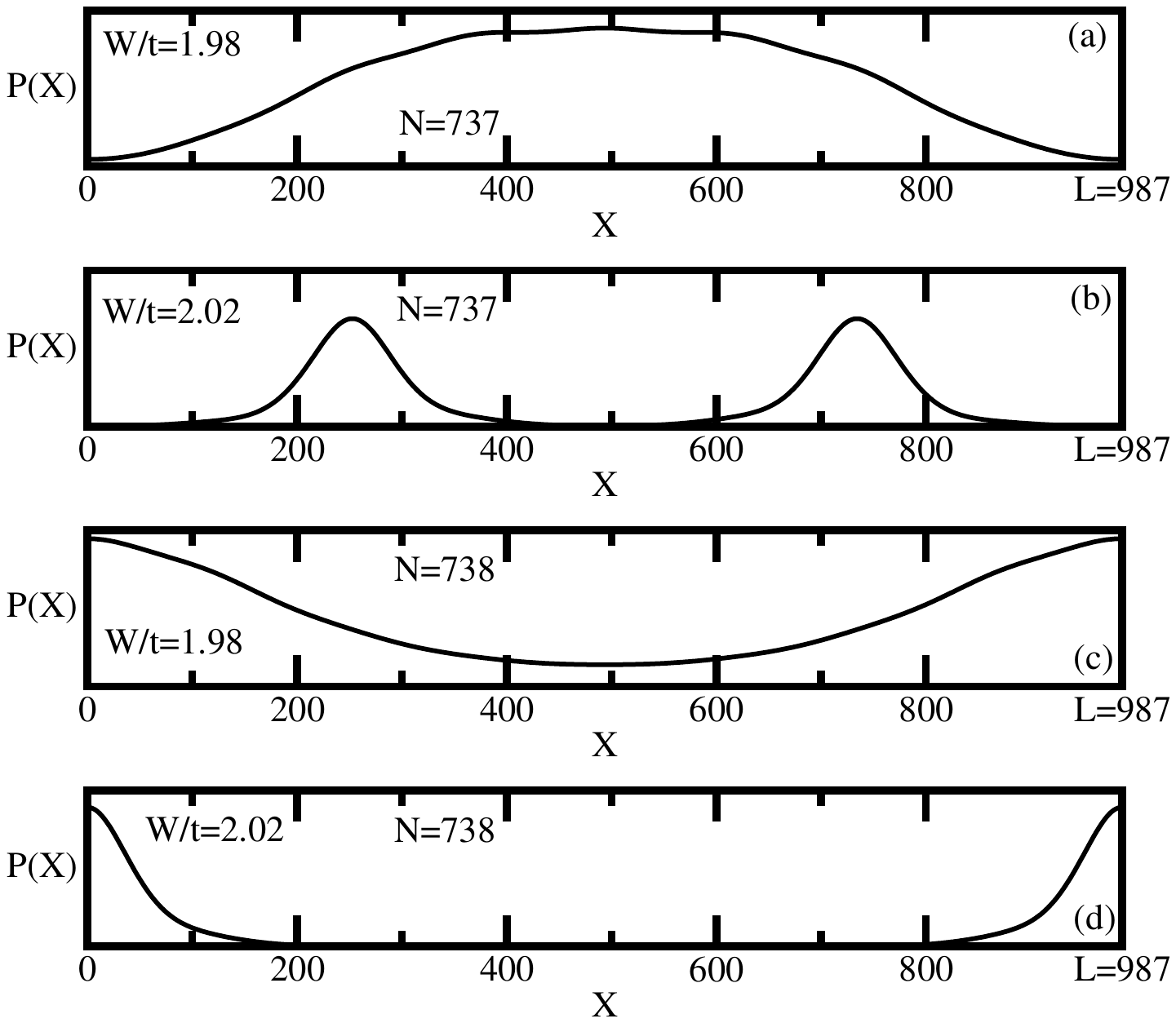}
 \caption{$P(X)$, distribution of the total position $X$, for four different calculations of system size $L=987$.  Panels (a) and (b) show the results for  $N=737$.   Panel (a)(panel (b)) shows results for $W/t=1.98$($W/t=2.02$).  Panels (c) and (d) show results for $N=738$.  Panel (c)(panel (d)) shows results for $W/t=1.98$($W/t=2.02$).}
 \label{fig:Px}
\end{figure}

Before presenting DLT results, we address a technical issue, that of odd/even $N$/$L$.  When $N$ and $L$ are either both even or odd, the distribution of the periodic total position $X$ (defined in Eq. (\ref{eqn:PX})) is bimodal near the critical $W/t$.   Fig. \ref{fig:Px} shows results for system size $L=987$, an odd system size.  Panels (a) and (b) show results for $N=737$ while panels (c) and (d) for $N=738$.   Panels (a) and (c) are for $W/t = 1.98$ (delocalized), panels (b) and (d) are for $W/t=2.02$, (localized).  In the localized phase, we see that the odd particle number system ($N=737$) shows two peaks, whereas the system with $N=738$ shows only one peak within one cell.  In other calculations  we always find that if the system size and particle number are either both even or both odd, the distribution in the localized phase is bimodal.  In this case the GBC formalism ($U_4$) is not applicable.  However, in the remaining cases (odd $N$ even $L$ and vice versa) the distribution is unimodal, and the GBC is applicable.   Even though the GBC (as derived in this work) is not applicable to the even/even and odd/odd cases, it is possible to determine the DLT critical point by direct inspection of $P(X)$.  This is somewhat laborious, so we focus on systems in which $N$ and  $L$ are of different parity.  We also carried out calculations in which the parities are the same and found that the all our conclusions stated below are maintained.\\

\subsection{Phase diagram and spikes} 

\label{ssec:GBC}

In the following, it is important to mention, that it is not possible to directly study irrational particle fillings numerically.  However, it is possible to study filling ratios for different system sizes which tend to specific irrational numbers as the system size is increased.  For example, the filling $N/L = F_{n-1}/F_n$ tends to $(\sqrt{5}-1)/2$.  It is also, of course, possible to study rational fillings by keeping $N/L$ a fixed ratio for different system sizes.  Below we will make comparisons between these two types of extrapolations. \\

The phase diagram obtained from our GBC calculations is shown in Fig. \ref{fig:pdfill}, in the main panel of that figure, for four system sizes, $L=377,987,1597,4181$, all odd Fibonacci numbers.  The scan in particle number includes all even $N$.  The phase diagram shows many examples of the predicted $W/t=2$, but spikes are also seen, for which $W/t<2$.  At these spikes $W/t$ is seen to decrease ($W \rightarrow 0$) with increasing system sizes.   \\

To investigate further we considered the three spikes in Fig. \ref{fig:pdfill}a which we labelled I, II, and III.  For I, we found that the filling can be written as $N/L = 2F_{n-2}/F_n = (F_{n-1} + F_{n-4})/F_n$.  As the system size is increased (as $n\rightarrow \infty$) $N/L (\rightarrow  3 - \sqrt{5})$.  In Fig. \ref{fig:pdfill}b the critical value of the interaction strength is shown as a function of system size, for $L=377,987,1597,4181,6765,17711$, covering almost two orders of magnitude.  In Figs. \ref{fig:pdfill}b, \ref{fig:pdfill}c, \ref{fig:pdfill}d, all axes are logarithmic.  The three plots are all well approximated by straight lines, suggesting that the critical potential strength decreases to zero in the thermodynamic limit according to a power law.  We fitted the function $f(L) = aL^{-b}$ to the data shown in Fig. \ref{fig:pdfill}a and found $a = 15.3(4)$ and $b = 0.479(4)$.   The spikes indicated by labels II and III (results shown in Figs \ref{fig:pdfill}c and Fig. \ref{fig:pdfill}d) exhibit the same features.  For spike II $N/L = \lim_{n \rightarrow \infty} \frac{F_{n-2} + F_{n-5}}{F_n} \rightarrow  2\sqrt{5} - 4$, $a = 7.6(1), b = 0.257(2)$, for spike III $N/L = \lim_{n \rightarrow \infty}(2F_{n-5}/F_n) \rightarrow 7 - 3\sqrt{5}$ and $a = 5.36(5)$ and $b = 0.177(1)$.  This same behavior, also hold for other spikes we checked: as the system size ($L=F_n$) is increased the fillings tend to a definite irrational number which is irrational since its expression includes the number $\sqrt{5}$.

In our actual calculation, all fillings are rational, of course, but some are better approximations to a given irrational number than others.   Suppose that one starts with a system size $L=F_n$, consider the filling $F_1/F_n$, which exhibits a transition at $W/t=2$,  being the worst approximation possible for the irrational number $\phi = \lim_{m \rightarrow \infty} F_{1+m}/F_{n+m}$, since $m=0$.   Increasing $L$ amounts to increasing $m$, as well as both $F_{1+m}$ and $F_{n+m}$, and the irrational limit $\phi$ will be better approximated.   As this happens, the transition approaches $W/t \rightarrow 0$.   At the same time, with every increased system $L$ or $m$, there will appear new "bad" rational approximations for other irrational fillings.  For example, for $m=1$, where the system size is $L=F_{n+1}$, there will be a new filling, $F_1/F_{n+1}$ which did not exist for $m=0$.  This is a new "bad" approximation for the irrational filling, $\phi' = \lim_{m \rightarrow \infty} F_{1+m}/F_{n+m+1}$, for which $W/t=2$.  Again, increasing $m$ and $L$ further will always improve the approximation for the irrational number to which the fillings at a given system size tend to, at the same time, larger $L$ will give rise to new fillings which are "bad" approximations to irrational numbers.  A similar, but complementary, effect can be expected to occur on the insulating side due to AAM duality~\cite{Johansson91} with momentum space localization~\cite{Hetenyi12,Hetenyi13} of many-particle states which consists of single-particle states which are delocalized in momentum space.\\

\begin{figure}[ht]
 \centering
 \includegraphics[width=7cm,keepaspectratio=true]{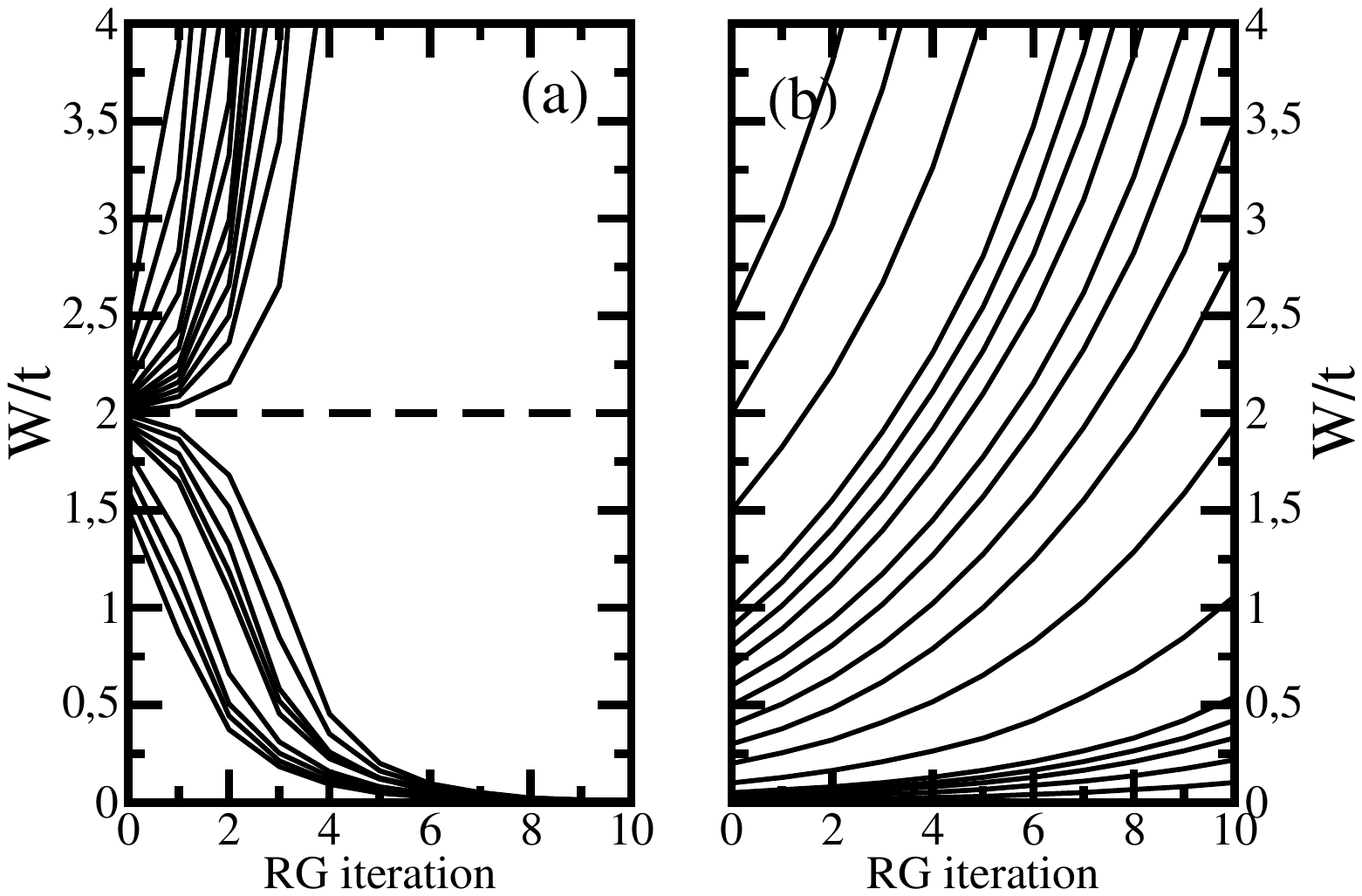}
 \caption{RG flow lines.  Panel (a) shows a calculation for half filling.  The iterations were carried out between $L= 2584$ and $L=610$.  The dashed line ($W = 2t$) indicates a repulsive fixed point.  The flowlines originating below(above) $W=2t$ tend to the attractive fixed point at $W = 0$($W \rightarrow \infty$).  Panel (b) shows a calculation for an irrational filling ($\lim_{n \rightarrow \infty} 2F_{n-2}/F_n$).   The iterations were carried out between the two systems $L=1597$, $N=1220$ and $L=987$, $N=754$.   In this case $W=0$($W \rightarrow \infty$) is a(n) repulsive(attractive) fixed point.  Flow lines started at finite $W$ all tend to infinity, indicating localization.}
 \label{fig:rgflow}
\end{figure}

\subsection{Polarization amplitude renormalization}  

\label{ssec:PAR}

\begin{figure}[ht]
 \centering
 \includegraphics[width=8cm,keepaspectratio=true]{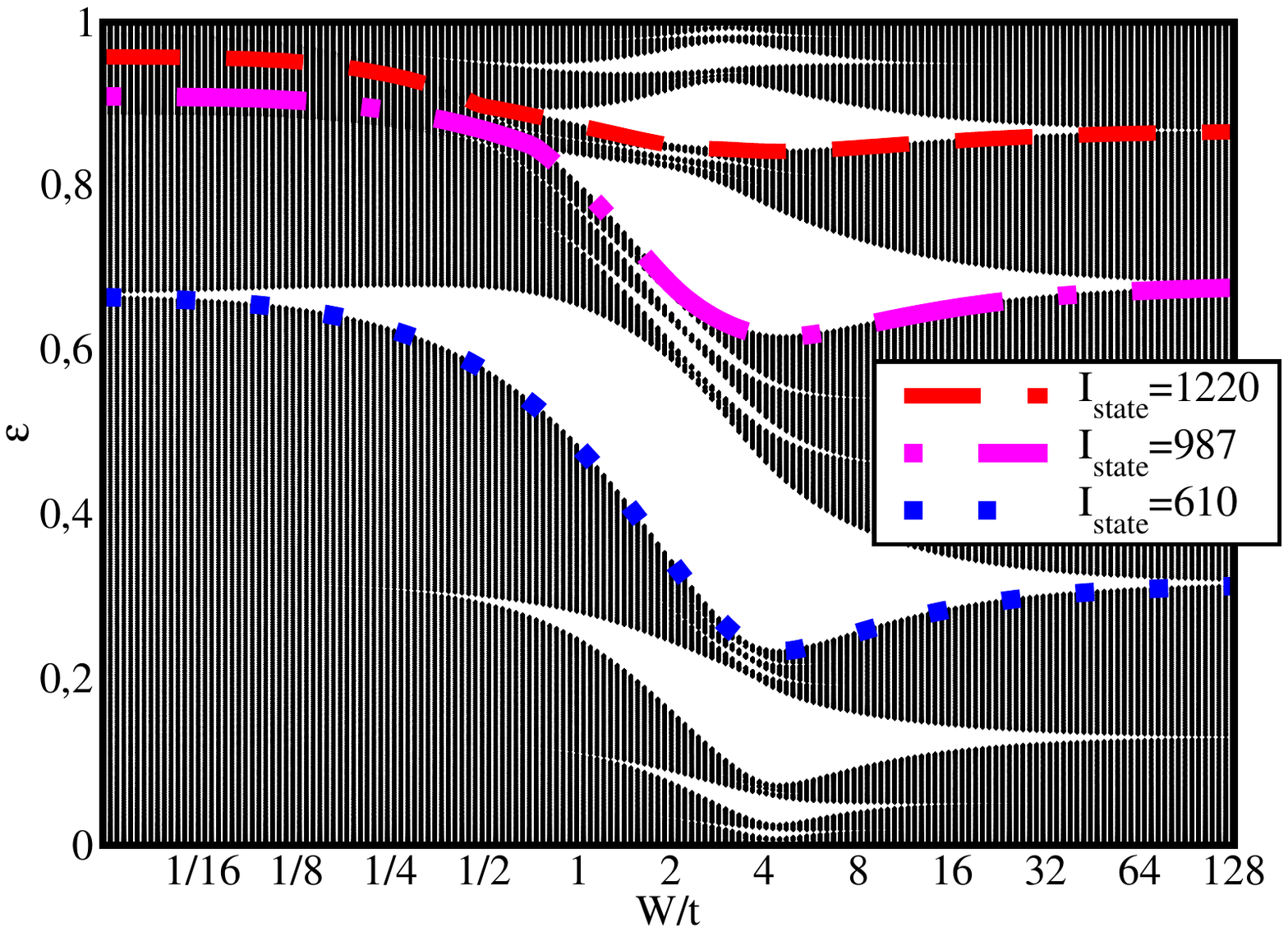}
 \caption{ Energy levels of the extended Aubry-Andr\'{e} model for a system of size $L=1597$ with periodic boundary conditions as a function of $W/t$.   The $W/t$-axis is plotted on a logarithmic scale.  $\epsilon$ on the vertical axis refers to a scaled and normalized energy eigenvalue according to $\epsilon = (E-E_{min})/(E_{max} - E_{min})$.  The energy levels of three particular states, with numbers $I_{state} = 1220, 987, 610$ are plotted as thick lines.  These filling numbers correspond to spikes I, II, and III in Fig. \ref{fig:pdt2rat}.}
 \label{fig:band_structure_t2}
\end{figure}

\begin{figure}[ht]
 \centering
 \includegraphics[width=8cm,keepaspectratio=true]{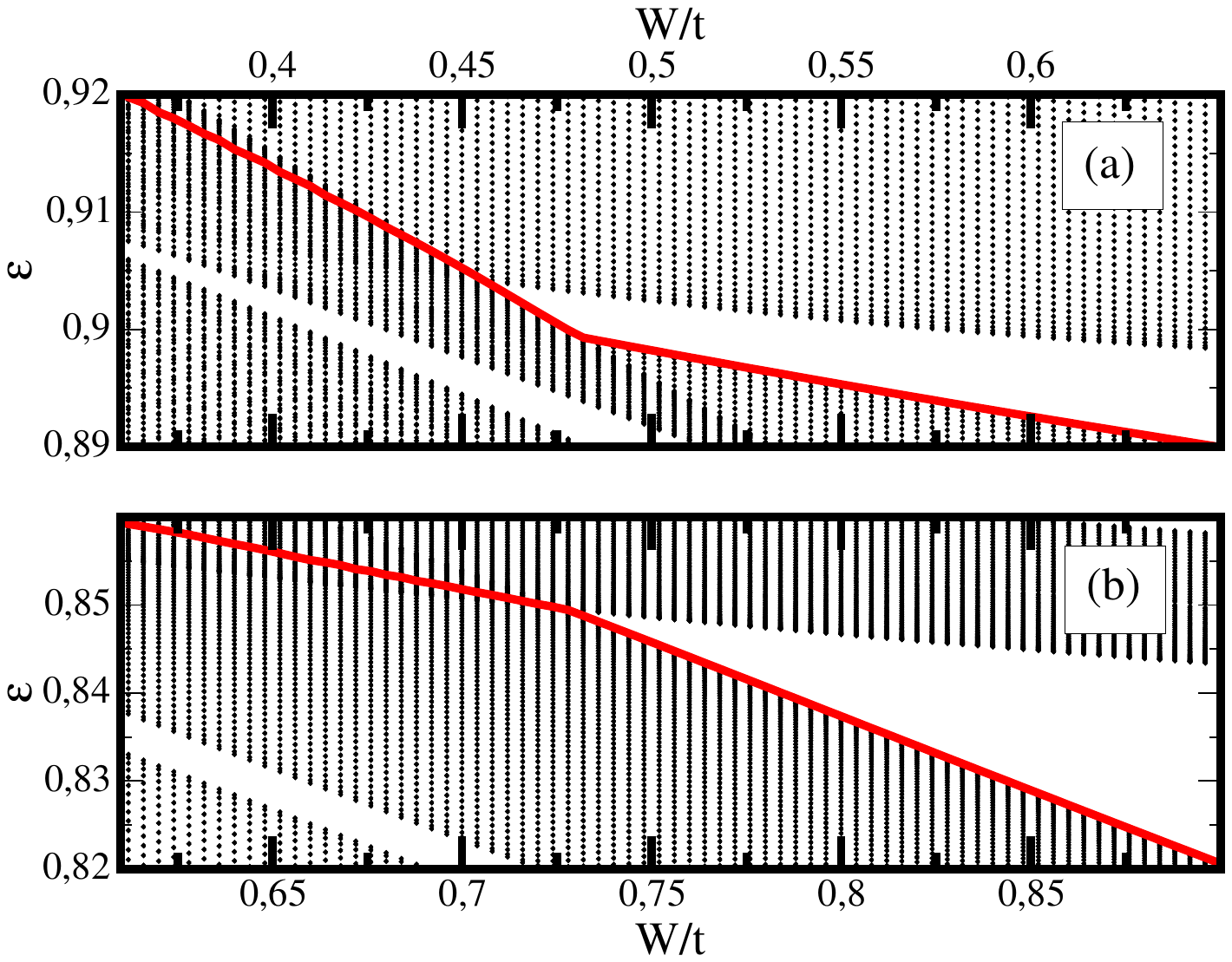}
 \caption{Zoomed band structure for an extended Aubry-Andr\'{e} model for a system of size $L=1597$ with periodic boundary conditions as a function of $W/t$.  The red solid lines in both panels represent states numbered $I_{state} = 1220$ (panel (a)) and $I_{state} = 987$ (panel (b)).  One can see clearly that these states correspond to gapless regions for smaller $W/t$ and gaps open at fintie $W/t$.  For $i_{state} = 1220$ the gap opening occurs at $W/t \approx 0.45$.  For $I_{state}=987$ the gap opening occurs at $W/t \approx 0.72$.}
 \label{fig:t2gaps}
\end{figure}

\begin{figure}[ht]
 \centering
 \includegraphics[width=7cm,keepaspectratio=true]{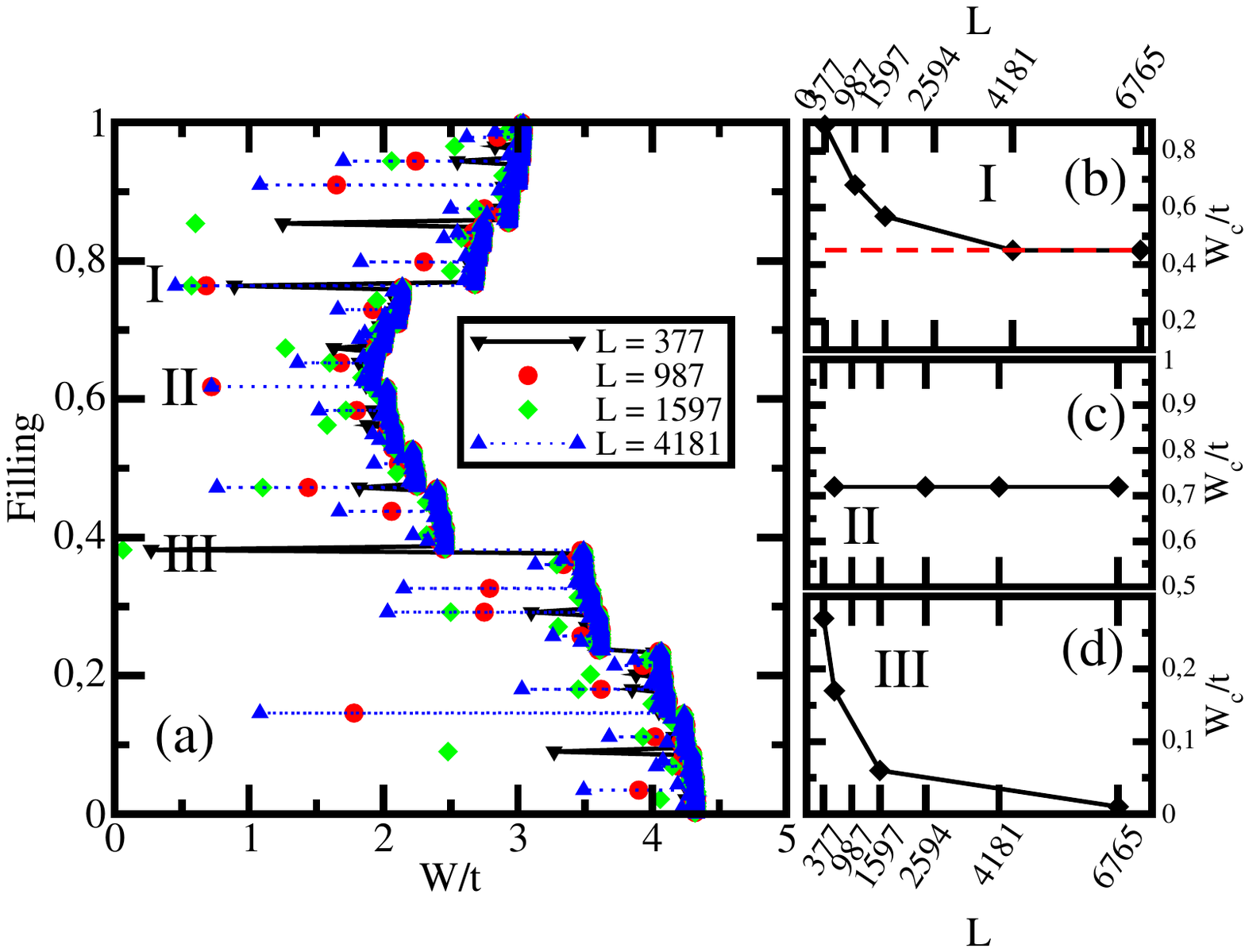}
 \caption{Panel (a): Phase diagram of the extended Aubry-Andr\'{e} model, with second nearest neighbor hopping with strength $t_2 = 0.5$.  Calculations based on four system sizes ($L=377,987,1597,4181$) are shown.  Fillings with even particle number were scanned.   Panel (b), (c), and (d): critical $W_c/t$ for the spike indicated in panel (a) by the Roman numeral I,II, and III, respectively.}
 \label{fig:pdt2rat}
\end{figure}

\begin{figure}[ht]
 \centering
 \includegraphics[width=7cm,keepaspectratio=true]{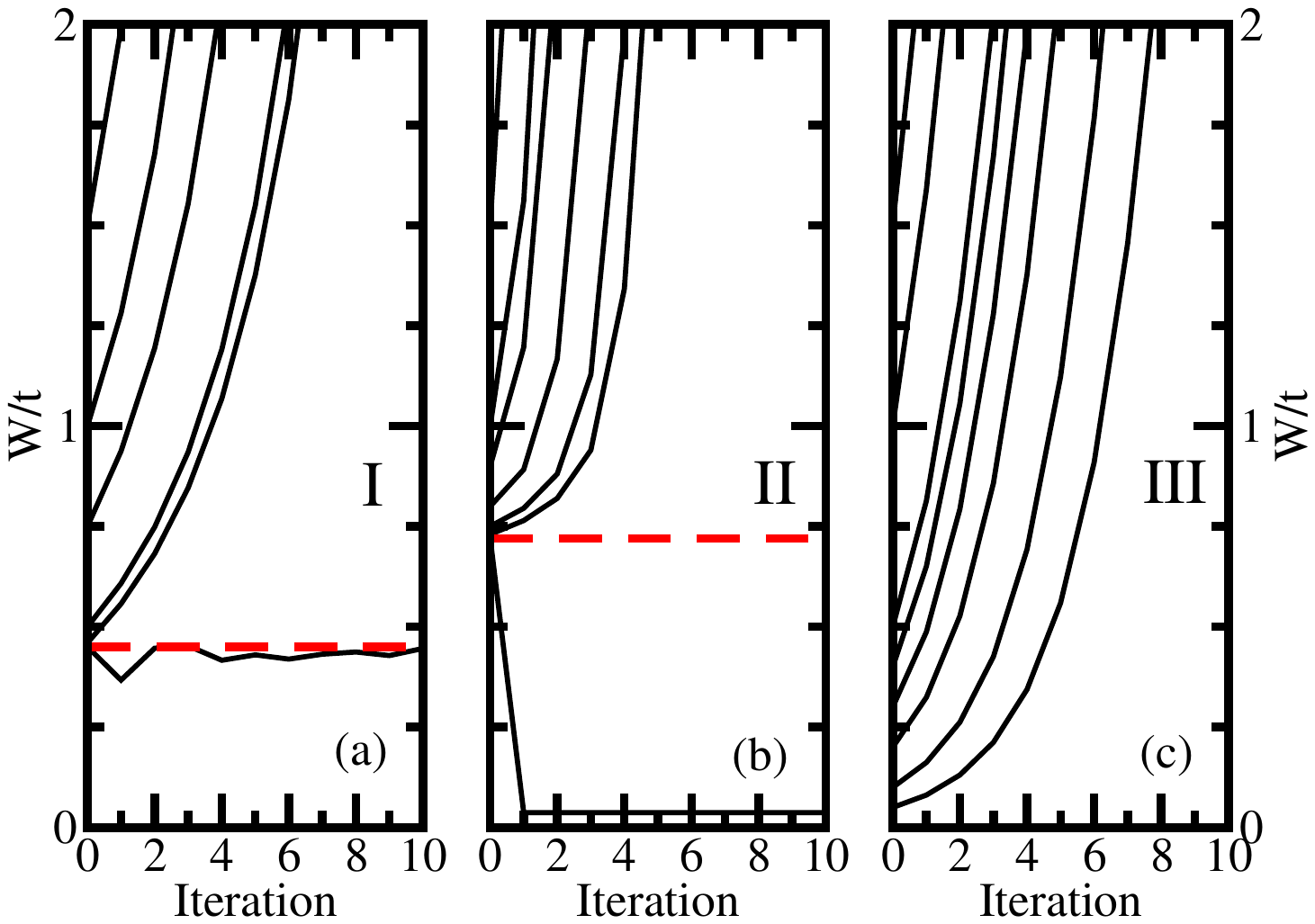}
 \caption{Renormalization flow lines for the three spikes indicated (I, II, III) in Fig. \ref{fig:pdt2rat}.  In all three cases the iterations took place by comparing between system sizes of $L'=1597$ and $L=987$ (Eq. (\ref{eqn:Z1RG})).  Panel (a) exhibits an attractive fixed point at $W \rightarrow \infty$.  Below $W/t<0.45$, where the transition is shown to occur in Fig. \ref{fig:pdt2rat}, the flow lines can not be calculated stably, because $Z_1\rightarrow 0$, but a repulsive fixed point is seen to occur at $W/t \approx 0.45$ (indicated with a red dashed line).  Panel (b) shows flow lines for  for spine II.  The behavior found is similar to spike I, except that the repulsive fixed point occurs at $W/t \approx 0.72$ (indicated with a red dashed line).  Panel (c) shows results for spike III, were a repulsive fixed point is seen to occur at $W/t = 0$, and an attractive one at $W/t \rightarrow \infty$.  All three results are consistent with the GBC calculations presented in Fig. \ref{fig:pdt2rat}. }
 \label{fig:flowt2}
\end{figure}

In Fig. \ref{fig:rgflow} two example PAR example calculations are shown, one for a filling of $1/2$ (rational), the other for a filling of $2 F_{n-2}/F_n \rightarrow 3 - \sqrt{5}$.  For the former, we used systems with $L=2584$ and $L=610$ (both even).  For the latter, we used $L=1597$ and $L=987$.   Fig. \ref{fig:rgflow}a shows a flow diagram exhibiting two attractive fixed points ($W=0$ and $W \rightarrow \infty$) and one repulsive fixed point at $W= 2t$.  The flow diagram for irrational filling (Fig. \ref{fig:rgflow}b) exhibits an attractive fixed point at $W \rightarrow \infty$ and a repulsive one at $W=0$. It is of interest to note that, while the GBC finds critical values of $W$ at finite values (which extrapolate to $W=0$ for large system sizes), the RG approach generates flow lines which immediately show that the localization transition occurs at $W=0$.   \\

\subsection{ Extended Aubry-Andr\'{e} model} 

For the extended AAM, defined in Eqs. (\ref{eqn:HAA}) and (\ref{eqn:Hext}), we repeat the same set of calculations done for the AAM above.  In this extended model, again we find "spikes", particular fillings at which the critical $W/t$ changes abruptly, but a key difference between this model and the AAM is that the spikes do not necessarily tend to zero.   The original AAM exhibits gaps which, as a function of $W/t$ open at $W/t=0$ and close at $W/t \rightarrow \infty$.  In the extended AAM we study, most gaps do not open at zero, but instead at some finite value, and the critcal $W/t$ at filled bands tends to the value where the gap opening occurs.  We present detailed results for this scenario. \\

Fig. \ref{fig:band_structure_t2} shows energy levels as a function of $W/t$ for a system of size $L=1597$.  The $W/t$ axis (horizontal) is logairthmic.  The energies are normalized and unitless, according to,
\begin{equation}
\epsilon = \frac{E_j - E_{min}}{E_{max} - E_{min}},
\end{equation}
for a given energy eigenvalue $E_j$.  An important difference between the extended model and the AAM is that in the former, some gaps open at $W/t=0$, but most gaps open at finite values of $W/t$, whereas in the AAM all gaps opened at $W/t=0$ and closed at $W/t \rightarrow \infty$.    Fig. \ref{fig:t2gaps} shows examples of zoomed band structure for two examples of finite gap openings for a system of $L=1597$.   Panel (a) (panel (b)) shows the case of $I_{state}=1220$($I_{state}=987$), which exhibit gap openings at $W/t\approx 0.45$($W/t=0.72$).  However, for the state  $I_{state}=610$ the gap opening and closing occur  at $W/t=0$ and $W/t \rightarrow \infty$, respectively, as can be seen in Fig. \ref{fig:band_structure_t2}.   The points where the gaps open will determine tha DLT, as argued below.\\

The DLT for the model given in Eq. (\ref{eqn:Hext}) is shown in Fig.  \ref{fig:pdt2rat} for odd system sizes ($L=377,987,1597,4181$), generated by scanning through even $N$s.   Spikes are found,  but in addition to Fig. \ref{fig:pdfill}, there are also "jumps" (horizontal) at irrational fillings.  Almost all such jumps are accompanied by spikes.   We can compare our results to two previous studies, Refs. \cite{Biddle11} and \cite{Varma15}.  Two phase diagrams are shown in Fig. 4 of Ref. \cite{Varma15}, albeit, ours is upside down compared to those, because our $t$ is negative.  The results shown in Ref. \cite{Varma15} are approximate, they show considerable error bars.  Comparing our results, however, to those of Ref. \cite{Biddle11} (also shown in Fig. 4 of Ref. \cite{Varma15}), we see that both exhibit discontinuities in the same places, although, ours are more pronounced, due to the increased precision.  The results of Ref. \cite{Biddle11} appear to be smeared out.\\

We also chose three different spikes, labeled by Roman numerals I, II, and III, in panel (a) to analyze, displayed in panels (b), (c), and (d), of Fig. \ref{fig:pdt2rat}, respectively.   The spikes correspond to the three fillings explicitly indicated in Fig. \ref{fig:band_structure_t2}, and their general formula are $2F_{n-2}/F_n = (F_{n-1} + F_{n-4})/F_n$, $F_{n-1}/F_n$, and $F_{n-2}/F_n$.  In these three cases, three different scenarios play out, as shown in these figure panels.  We determined that the gap in Fig. \ref{fig:band_structure_t2} for $N=1220$ and $L=1597$ opens at $W/t=0.45$.  In panel (b) of Fig. \ref{fig:pdt2rat} we see that the critical interaction strength occurs at $W/t > 0.45$ for smaller system sizes, but floors at this value as the system size is increased.  Panel (c) of the same figure, for fillings represented by $F_{n-1}/F_n$, shows all critical potential strengths $W/t = 0.72$, which is where the gap opening occurs (we determined this by zooming in the results of Fig. \ref{fig:band_structure_t2}).  The results for panel (d) are similar to the scenarios found for the standard AAM, the critical $W/t$ approaches zero.  For this filling the gap as a function of $W/t$ opens at $W=0$ (Fig. \ref{fig:band_structure_t2}).  These results coincide with the band structure calculations and gap openings shown in Fig. \ref{fig:t2gaps}. \\

Flow lines from the PAR calculatons for the three spikes I, II, and III are shown in Fig. \ref{fig:flowt2}.   In all three cases the iterations took place by comparing between system sizes of $L'=1597$ and $L=987$ (Eq. (\ref{eqn:Z1RG})).   Panel (a) presents the flow lines for spike I, for which a gap closure was found at $W/t \approx 0.45$.  Indeed, in this case, a repulsive fixed points is found there.  Flow lines which start at $W/t < 0.45$ are difficult to calculate.  In this case $|Z_1| \approx 0$ and the comparison between $Z_1$ for the different system sizes is numerically unstable.  Interesingly, in the standard AAM calculation, where the localization is not accompanied by gap closure, we did not encounter this issue.  Panel (b) shows the flow lines for spike II, where gap closure occurs at $W/t \approx 0.72$, which is where the repulsive fixed point is found to occur in this case, subject to the same numerical limitation as spike I (panel (a)).  Panel (c) shows the flow lines of spike III.  In this case a repulsive fixed point is found at $W/t = 0$, while an attractive one occurs at $W/t \rightarrow \infty$.  The results shown in Fig. \ref{fig:flowt2} (based on PAR calculations) are consistent with those shown in Fig. \ref{fig:pdt2rat} which are GBC results. \\

\section{Conclusion}  

\label{sec:cnclsn}

We investigated the Aubry-Andr\'{e} model, and one of its extensions.  In particular, we calculated the phase diagram as a function of particle density (filling) and interaction strength.  We find that upon extrapolating to the thermodynamic limit, the critical interaction strength depends on whether a given filling tends to a rational or an irrational number.  For the former, we find $W=2t$, a result expected from previous investigations, mostly on single particle states.  However, at irrational fillings, we find that the critical interaction strength tends to zero.  The irrational numbers accessible in our calculations are not the entire set, but only those, which can be produced as the limits of ratios of Fibonacci numbers (and sums thereof).  They are irrational because they all depend on the number $\sqrt{5}$.   Our calculations for an extended model, with second nearest neighbor hopping, also exhibits spikes at the same fillings as the original model, but the spikes have lower limits corresponding to gap closures, which in the extended model, occur at finite values of the potential strength.\\

Our work may have interesting implications for generalized AAMs as well as for quantum Hall studies, which, often, are based on the TKNN~\cite{Thouless82} invariant.  This formalism is usually restricted to the case of rational flux.   Due to the large change in localization properties between rational and irrational fillings, the AAM may serve as a template in the construction of  high-to-low conductivity switches, an effect predicted~\cite{Dey20} in small  AA rings.  Our methods are applicable to related disordered models in which the distinction between rational and irrational filling may be of interest (generalized AAM) or other quasiperiodic models such as the Fibonacci chain~\cite{Jagannathan21}.\\

\section{acknowledgments}  
This research was supported by HUN-REN 3410107 (HUN-REN-BME-BCE Quantum Technology Research Group), by the National Research, Development and Innovation Fund of Hungary within the Quantum Technology National Excellence Program (Project No. 2017-1.2.1-NKP-2017-00001), by Grants No. K142179 and No. K142652, and by the BME-Nanotechnology FIKP Grant No. (BME FIKP-NAT).  \\

\section*{Appendix: the Zekendorf representation of natural numbers as Fibonacci sums}

In this appendix we present the Zekendorf construction to find the sum of Fibonacci numbers to represent natural numbers.  The complete proof that this representation is unique, meaning each natural number is represented by a distinct sum (under the condition that no consecutive Fibonacci numbers appears in the sum), is given in Ref. \cite{Vajda89}.  The particular recipe we present below can be found in Ref. \cite{Samons94} \\

Let $N$ be a natural number.  One can find a Fibonacci number, $F_{m_1}$, for which it holds that
\begin{equation}
F_{m_1} \leq N < F_{m_1+1}.
\end{equation}
This also means, that
\begin{equation}
N - F_{m_1} < F_{m_1 + 1} - F_{m_1} = F_{m_1-1} < F_{m_1}.
\end{equation}
If $N= F_{m_1}$, then done.  If not, then let $N_1 = N - F_{m_1}$.  The same argument can now be applied to $N_1$, that is, one can find a Fibonacci number, $F_{m_2}$, such that
\begin{equation}
F_{m_2} \leq N_1 < F_{m_2+1}.
\end{equation}
This also means, that
\begin{equation}
N_1 - F_{m_2} < F_{m_2 + 1} - F_{m_2} = F_{m_2-1} < F_{m_2}.
\end{equation}
Carrying the above steps iteratively, at some point, one will encounter a Fibonacci number for some $N_k = F_{m_{k+1}}$.  Thus, we can write,
\begin{equation}
N = \sum_{w=1}^{k+1} F_{m_w}.
\end{equation}


\begin{thebibliography}{9} 

\bibitem{Abrahams79} E. Abrahams, P. W. Anderson, D. C. Licciardello, and T. V. Ramakrishnan, "Scaling Theory of Localization: Absence of Quantum Diffusion in Two Dimensions" {\it Phys. Rev. Lett.} {\bf 42} 673 (1979).

\bibitem{Langedijk09} A. Langedijk, B. van Tiggelen, and D. S. Wiersma, "Fifty years of Anderson localization" {\it Physics Today} {\bf 62} 24 (2009).

\bibitem{Evers08} F. Evers and A. D. Mirlin, "Anderson transitions" {\it Rev. Mod. Phys.} {\bf 80} 1355 (2008).
 
\bibitem{Aubry80} S. Aubry and G. Andr\'{e}, "Analyticity breaking and Anderson localization in incommensurate lattices." {\it Ann. Isr. Phys.} {\bf 3} 133 (1980).

\bibitem{Martinez18} A. J. Martinez, M. A. Porter, and P. T. Kevrekidis, "Quasiperiodic granular chains and Hofstadter butterflies" {\it Philos. Trans. A} {\bf 376} 20170139 (2018). 

\bibitem{Dominguez-Castro19} G. A. Dominguez-Castro, R. Paredes,  "The Aubry–André model as a hobbyhorse for understanding the localization phenomenon", {\it Eur. J. Phys.} {\bf 40} 045403 (2019).

\bibitem{Billy08} J. Billy, V. Josse, Z. Zuo, A. Bernard, B. Hambrecht, P. Logan, D. Cl\'{e}ment, L. Sanchez-Palencia, P. Bouyer, and A. Aspect, "Direct observation of Anderson localization of matter-waves in a controlled disorder" {\it Nature (London)} {\bf 453} 891 (2008).

\bibitem{Roati08} G. Roati, C. D'Errico, L. Fallani, M. Fattori, C. Fort, M. Zaccanti, G. Modugno, M. Modugno, and M. Inguscio, . "Anderson localization of a non-interacting Bose–Einstein condensate" {\it Nature (London)} {\bf 453} 895 (2008).

\bibitem{Modugno09} M. Modugno, "Exponential localization in one-dimensional quasi-periodic optical lattices" {\it New. J. Phys.} {\bf 11} 033023 (2009).

 \bibitem{Johansson91} M. Johansson and R. Riklund, "Self-dual model for one-dimensional incommensurate crystals including next-nearest-neighbor hopping, and its relation to the Hofstadter model" {\it Phys. Rev. B} {\bf 43} 13468 (1991).

 \bibitem{Harper55} P. G. Harper, "Single Band Motion of Conduction Electrons in a Uniform Magnetic Field"{\it Proc. Phys. Soc. A} {\bf 68} 874 (1955).
 
 \bibitem{vonKlitzing80} K. von Klitzing, G. Dorda, and M. Pepper, "New Method for High-Accuracy Determination of the Fine-Structure Constant Based on Quantized Hall Resistance" {\it Phys. Rev. Lett.} {\bf 45} 494 (1980).
 
 \bibitem{Tsui82} D. C. Tsui, H. L. Stormer, and A. C. Gossard, "Two-Dimensional Magnetotransport in the Extreme Quantum Limit"{\it Phys. Rev. Lett.} {\bf 48} 1559 (1982).
 
  \bibitem{Thouless82} D. J. Thouless, M. Kohmoto, M. P. Nightingale, M. den Nijs, "Quantized Hall Conductance in a Two-Dimensional Periodic Potential" {\it Phys. Rev. Lett.} {\bf 49 } 405 (1982).
  
\bibitem{Biddle10} J. Biddle and S. Das Sarma, "Predicted Mobility Edges in One-Dimensional Incommensurate Optical Lattices: An Exactly Solvable Model of Anderson Localization"  {\it Phys. Rev. Lett.} {\bf 104} 070601 (2010).

 \bibitem{Biddle11}  J. Biddle, D. J. Priour Jr., B. Wang, and S. Das Sarma, "Localization in one-dimensional lattices with non-nearest-neighbor hopping: Generalized Anderson and Aubry-André models" {\it Phys. Rev. B} {\bf 83} 075105 (2011).
  
 \bibitem{Ganeshan13} S. Ganeshan, K. Sun, and S. Das Sarma, "Topological Zero-Energy Modes in Gapless Commensurate Aubry-André-Harper Models"  {\it Phys. Rev. Lett.} {\bf 110} 180403 (2013).
 
   \bibitem{Ganeshan15} S. Ganeshan, J. H. Pixley, and S. Das Sarma, "Nearest Neighbor Tight Binding Models with an Exact Mobility Edge in One Dimension" {\it Phys. Rev. Lett.} {\bf 114} 146601 (2015).
 
 \bibitem{Bistritzer11} R. Bistritzer and A. H. Macdonald, "Moiré bands in twisted double-layer graphene", {\it Proc. Nat. Acad. Sci. USA} {\bf 108} 12233 (2011).

\bibitem{Monthus17} C. Monthus, "Multifractality in the generalized Aubry-Andre quasiperiodic localization
model with power-law hoppings or power-law Fourier coefficients" {\it Fractals} {\bf 27} 1950007 (2017).

 \bibitem{Li20} X. Li and S. Das Sarma, "Mobility edge and intermediate phase in one-dimensional incommensurate lattice potentials", {\it Phys. Rev. B} {\bf 101} 064203 (2020).

\bibitem{Padhan22} A. Padhan, M. K. Giri, S. Mondal, and T. Mishra, "Emergence of multiple localization transitions in a one-dimensional quasiperiodic lattice" {\it Phys. Rev. B} {\bf 105} L220201 (2022).
 
\bibitem{Goncalves23a} M. Gon\c{c}alves, B. Amorim, E. V. Castro, and P. Ribeiro, "Critical Phase Dualities in 1D Exactly Solvable Quasiperiodic Models" {\it Phys. Rev. Lett.} {\bf 131} 186303 (2023).

\bibitem{Dziurawiec24} M. Dziurawiec, J. O. de Almeida, M. L. Bera, M. P\/{l}odzie\'{n}, M. M. Maśka, M. Lewenstein, T. Grass, and U. Bhattacharya, "Unraveling multifractality and mobility edges in quasiperiodic Aubry-André-Harper chains through high-harmonic generation" {\it Phys. Rev. B} {\bf 110} 014209 (2024).

\bibitem{Goncalves23b} M. Gon\c{c}alves, B. Amorim, E. V. Castro, and P. Ribeiro, "Renormalization group theory of one-dimensional quasiperiodic lattice models with commensurate approximants" {\it Phys. Rev. B} {\bf 108} L100201 (2023).

 \bibitem{Papp07} E. Papp and C. Micu, {\it Low-Dimensional Nanoscale Systems on Discrete Spaces}, World Scientific, Singapore, (2007).

\bibitem{Dey20} S. Dey, D. Daw, S. K. Maiti, "Flux-driven circular current and near-zero field magnetic response in an Aubry ring: High-to-low conducting switching action" {\it EPL} {\bf 129} 47002 (2020).

\bibitem{Ganguly23} S. Ganguly and S. K. Maiti "Electrical analogue of one-dimensional and quasi-one-dimensional Aubry–André–Harper lattices" {\it Sci. Rep.} {\bf 13} 13633 (2023).

\bibitem{Jitomirskaya99} S. Ya. Jitomirskaya,  "Metal-Insulator Transition for the Almost Mathieu Operator" {\it Ann. Math.} {\bf 150}  1159 (1999).

\bibitem{Avila09} A. Avila and S. Jitomirskaya "The Ten Martini Problem" {\it Ann. Math.} {\bf 170} 303 (2009).

\bibitem{Avila23} A. Avila, J. You, and Q. Zhou, "Dry Ten Martini Problem in the non-critical case" arxiv:2306.16254. 

\bibitem{Wang17} Y. Wang, G. Xianlong, and S. Chen, "Almost mobility edges and the existence of critical regions in one-dimensional quasiperiodic lattices" {\it Eur. Phys. J. B} {\bf 90} 215 (2017).

\bibitem{Zhang15} Y. Zhang, D. Bulmash, A. V.  Maharaj, C.-M. Jian, and S. A. Kivelson, "The almost mobility edge in the almost Mathieu equation" arXiv:1504.05205.

\bibitem{Hetenyi19} B. Het\'{e}nyi and B. D\'{o}ra, "Quantum phase transitions from analysis of the polarization amplitude" {\it Phys. Rev. B} {\bf 99} 085126 (2019).

\bibitem{Hetenyi22} B. Het\'{e}nyi and S. Cengiz, "Geometric cumulants associated with adiabatic cycles crossing degeneracy points: Application to finite size scaling of metal-insulator transitions in crystalline electronic systems" {\it Phys. Rev. B} {\bf 106}  195151 (2022).

 \bibitem{Hetenyi24} B. Het\'{e}nyi, "Scaling of the bulk polarization in extended and localized phases of a quasiperiodic model" {\it Phys. Rev. B} {\bf 110} 124125 (2024).
 
 \bibitem{Binder81a} K. Binder, "Finite size scaling analysis of ising model block distribution functions" {\it Z. Phys. B } {\bf 43} 119 (1981).
  
\bibitem{Binder81b} K. Binder,  "Critical Properties from Monte Carlo Coarse Graining and Renormalization" {\it Phys. Rev. Lett. } {\bf 47} 693 (1981).

\bibitem{Vanderbilt18} D. Vanderbilt, {\it Berry Phases in Electronic
  Structure Theory}, Cambridge University Press, Cambridge,
  U.K. (2018).
  
\bibitem{Resta00} R. Resta, "Manifestations of Berry's phase in molecules and condensed matter" {\it J. Phys.: Cond. Mat.} {\bf 12} R107
  (2000).
  
  \bibitem{King-Smith93} R. D. King-Smith and D. Vanderbilt, "Theory of polarization of crystalline solids" {\it
  Phys. Rev. B} {\bf 47} R1651 (1993).

\bibitem{Resta94} R. Resta, "Macroscopic polarization in crystalline dielectrics: the geometric phase approach" {\it Rev. Mod. Phys.} {\bf 66} 899
  (1994).
  
  \bibitem{Hetenyi21} B. Het\'{e}nyi, S. Parlak, and M. Yahyavi, "Scaling and renormalization in the modern theory of polarization: Application to disordered systems" {\it Phys. Rev. B} {\bf 104} 214207 (2021).

\bibitem{Kohn64} W. Kohn, {\it Phys. Rev.} "Theory of the Insulating State" {\bf 133} A171 (1964).

 \bibitem{Bernevig13} B. A. Bernevig and T. L. Hughes, {\it Topological Insulators and Superconductors}, Princeton University Press (2013).

\bibitem{Asboth16} J. K. Asb\'{o}th, L. Oroszl\'{a}ny, and A. P\'{a}lyi, {\it A Short Course on Topological Insulators: Band Structure and Edge States in One and Two Dimensions}, Lecture Notes on Physica, vol. 919, Springer International Publishing, (2016).

 \bibitem{Kitaev01} A. Yu. Kitaev, "Unpaired Majorana fermions in quantum wires" {\it Phys.-Usp.} {\bf 44} 131 (2001).
 
  \bibitem{Kobayashi18} R. Kobayashi, Y. O. Nakagawa, Y. Fukusumi, and M. Oshikawa, "Scaling of the polarization amplitude in quantum many-body systems in one dimension" {\it Phys. Rev. B} {\bf 97} 165133 (2018).
 
 \bibitem{Wilkinson84} M. Wilkinson, "Critical properties of electron eigenstates in incommensurate systems" {\it Proc. Roy. Soc. Lond. A} {\bf 391} 305 (1984).
 
 \bibitem{Niu86} Q. Niu and F. Nori, "Renormalization-Group Study of One Dimensional Quasiperiodic Systems", {\it Phys. Rev. Lett.} {\bf 57}, 2057
(1986).

\bibitem{Ashraff88}  J. A. Ashraff and R. B. Stinchcombe, "Exact decimation approach to the Green’s functions of the Fibonacci-chain quasicrystal", {\it Phys. Rev. B} {\bf 37}, 5723 (1988).

 \bibitem{Andrews24} B. Andrews, D. Reiss, F. Harper, and R. Roy, "Localization renormalization and quantum Hall systems" {\it Phys. Rev. B} {\bf 109} 125132 (2024).
 
 \bibitem{Vajda89} S. Vajda, "Fibonacci and Lucas Numbers, and the Golden Section: Theory and Applications", Ellis Horwood Limited, Chichester, U.K., 1989.
 
 \bibitem{Samons94}  J. D. Samons, "A Relationship Between the Fibonacci Sequence and Cantor's Ternary Set", UNF Graduate Theses and Dissertations. 285, (1994).  (https://digitalcommons.unf.edu/etd/285).
 
 \bibitem{Hetenyi12} B. Het\'enyi, "Current Response in Extended Systems as a Geometric Phase: Application to Variational Wavefunctions", {\it J. Phys. Soc. Jpn.} {\bf 81} 124711 (2012).
 
 \bibitem{Hetenyi13} B. Het\'enyi, "dc conductivity as a geometric phase" {\it Phys. Rev. B} {\bf 87} 235123 (2013).

\bibitem{Varma15} V. Kerala Varma and S. Pilati, "Kohn's localization in disordered fermionic systems with and without interactions" {\it Phys. Rev. B} {\bf 92} 134207 (2015).

 \bibitem{Jagannathan21} A. Jagannathan, "The Fibonacci quasicrystal: Case study of hidden dimensions and multifractality" {\it Rev. Mod. Phys.} {\bf 93} 045001 (2021).
 
 \end{thebibliography}
\end{document}